# Overcoming the surface paradox: Buried perovskite quantum dots in wide-bandgap perovskite thin films

Hao Zhang[1,2], Altaf Pasha[1,3#], Isaac Metcalf[4#], Jianlin Zhou[1], Mathias Staunstrup[5,6], Yunxuan Zhu[7], Shusen Liao[2], Ken Ssennyimba[2], Jia-Shiang Chen[8], Surya Prakash Reddy[9], Simon Thébaud[10], Jin Hou[4], Xinting Shuai[4], Faiz Mandani[1], Siraj Sidhik[1,4], Matthew R. Jones[11], Xuedan Ma[4,8], R Geetha Balakrishna[3], Sandhya Susarla[9], David S. Ginger[12], Claudine Katan[13], Mercouri G. Kanatzidis[14], Moungi G. Bawendi[15], Douglas Natelson[4,7,16], Philippe Tamarat[5,6], Brahim Lounis[5,6], Jacky Even[10*] and Aditya D. Mohite[1,4*]

**[1]Department of Chemical and Biomolecular Engineering, Rice University, Houston, Texas 77005, USA.**

**[2]Applied Physics Program, Smalley-Curl Institute, Rice University, Houston, TX, 77005, USA.**

**[3]Centre for Nano and Material Sciences, Jain University, Jain Global Campus, Kanakapura, Bangalore – 562112, Karnataka, India.**

**[4]Department of Materials Science and NanoEngineering, Rice University, Houston, TX, 77005, USA.**

**[5]Université de Bordeaux, LP2N, Talence, France.**

**[6]Institut d'Optique and CNRS, LP2N, Talence, France.**

**[7]Department of Physics and Astronomy, Rice University, Houston, Texas 77005, USA.**

**[8]Center for Nanoscale Materials, Argonne National Laboratory, Lemont, IL 60439, USA.**

**[9]School for Engineering of Matter, Transport and Energy, Arizona State University, Tempe, AZ, 85281, USA.**

**[10]Univ Rennes, CNRS, Institut FOTON (Fonctions Optiques pour les Technologies de l'Information), UMR 6082, CNRS, INSA de Rennes, 35708 Rennes, France.**

**[11]Department of Chemistry, Rice University, Houston, TX, 77005, USA.**

**[12]Department of Chemistry, University of Washington, Box 351700, Seattle, Washington 98195-1700, USA.**

**[13]Univ Rennes, CNRS, ISCR (Institut des Sciences Chimiques de Rennes), UMR 6226, CNRS, 35000 Rennes, France.**

**[14]Department of Chemistry, Northwestern University, Evanston, IL, USA.**

**[15]Department of Chemistry, Massachusetts Institute of Technology, 77 Massachusetts Avenue, Cambridge, MA 02139, USA.**

[16]Department of Electrical and Computer Engineering, Rice University, Houston, Texas 77005, USA.

*Correspondence: jacky.even@insa-rennes.fr & adm4@rice.edu

[#]These authors contributed equally to this work.

**Colloidal perovskite quantum dots (PQDs) are an exciting platform for on-demand quantum, and classical optoelectronic and photonic devices. However, their potential success is limited by the extreme sensitivity and low stability arising from their weak intrinsic lattice bond energy and complex surface chemistry. Here we report a novel platform of buried perovskite quantum dots (b-PQDs) in a three-dimensional perovskite thin-film, fabricated using one-step, flash annealing, which overcomes surface related instabilities in colloidal perovskite dots. The b-PQDs demonstrate ultrabright and stable single-dot emission, with resolution-limited linewidths below 130 μeV, photon-antibunching ($g^2(0)$=0.1), no blinking, suppressed spectral diffusion, and high photon count rates of $10^4$/s, consistent with unity quantum yield. The ultrasharp linewidth resolves exciton fine-structures (dark and triplet excitons) and their dynamics under a magnetic field. Additionally, b-PQDs can be electrically driven to emit single photons with 1 meV linewidth and photon-antibunching ($g^2(0)$=0.4). These results pave the way for on-chip, low-cost single-photon sources for next generation quantum optical communication and sensing.**

Halide perovskite quantum dots (or nanocrystals) with a general formula of $AMX_3$ (A = Cs, MA, FA (FA, formamidinium; MA, methylammonium); B=Pb, Sn, Ge and X = Cl, Br, I) have emerged as an exciting platform for building the next generation of integrated classical and quantum emitters integrated with photonic circuits.[1,2] Moreover, they offer an opportunity to simultaneously achieve optically or electrically driven quantum emitters with high brightness, indistinguishability, photon purity, using deterministic and scalable fabrication scheme, which has been only possible in III-V quantum dots on micropilar cavities, grown using molecular beam epitaxy.[3] However, achieving these properties in solid-state materials is challenging due to interactions with phonons, dark states, impurities, and dielectric fluctuations, causing spectral diffusion over tens to hundreds of microseconds.[4–6] Many efforts have been made to resolve the stability of quantum dots, through the design of new ligands chemistries, and surface passivation techniques, such as growing core-shell structures.[7–10] More recently, new strategies using phospholipid capping ligands strategies have been shown to enhance the structural and colloidal integrity.[9] However, strategies relying on building core-shell architectures have proved to be challenging, unlike the II-VI colloidal quantum dots. Resolving these surface-related instabilities has been the bane of colloidal PQDs over the last decade and critical for unlocking their potential for applications.[8] As a result, the near-ideal properties observed in freshly synthesized colloidal perovskite quantum dots (c-PQDs) are transient in nature and degrade rapidly during their interrogation. In addition, several of the ligand strategies are not compatible with charge transport and injection for the development of electrically driven optoelectronic devices.[1,11] Here we show that we can resolve this long-enduring challenge of surface chemistry by introducing a new material platform of buried perovskite quantum dots (b-PQDs), which pave the path for building solution-processable, integrated quantum and classical optoelectronic and photonic devices across the visible and near IR regimes.

## Growth of buried perovskite quantum dots (b-PQDs)

The b-PQDs in the host perovskites were formed using a one-step synthesis approach during thin film formation. As described in **Fig. 1a**, a precursor solution for 3D perovskite formamidinium lead bromide ($FAPbBr_3$) was doped with iondine additives as nucleation seeds, followed by sequential annealing process. Cross-sectional transmission electron microscopy (TEM) was conducted to characterize the local crystalline microstructure and composition of the b-PQDs. **Fig. 1b** and **1c** show TEM images on the synthesized films (with additive), indicating the presence of nanometer-sized domains, with a heterogenous size distribution of the embedded PQDs (3 – 10 nm, **Supplementary Fig. 1**). The b-PQD displayed in **Fig. 1c** reveals consistent spacing of lattice fringes of 2.8 ± 0.2 Å, corresponding to the (210) planes from a $FAPb(I_{1-x}Br_x)_3$ phase (**Fig. 1c**). In contrast, a uniform landscape and consistent atomic arrangement are detected in undoped $FAPbBr_3$, suggesting a single phase of the pristine perovskite host (**Supplementary Fig. 1a**). An in-depth analysis of the b-PQDs composition based on FFT is shown in **Supplementary Fig. 1d - 1f**. $FAPbBr_3$ and $FAPbI_3$ have distinct lattice parameters that can potentially be distinguished from one another using the diffraction patterns. As shown in **Supplementary Fig. 1f**, the Fast Fourier Transform (FFT) from the entire image was categorized into three sets of distinct FFTs. Eventually, all the three sets were FFT filtered and combined to produce a composite RGB image as shown in **Supplementary Fig. 1e**. The red and blue regions represent a Br-rich phase (lattice parameter of 5.8 ± 0.2 Å). The green region represents I-rich phase with a lattice parameter of 6.2 ± 0.2 Å. It can be observed that most of the quantum dot-like phase is dominant in iodine composition. Lattice parameters reported in the literature for both pure and unstrained $FAPbBr_3$ (a = 5.986 Å)[12] and $FAPbI_3$ (a = 6.365 Å)[13,14] do not match exactly with the observed lattice parameters, but the lattice mismatch between the two pure compounds is reflected by the local variations of lattice parameters in the TEM images. Strain or compositional I/Br mixing may also be present in the system.

Next, we studied the optical properties of the films containing the b-PQDs. The ensemble optical absorbance and photoluminescence (PL) spectra of the synthesized b-PQDs system are shown in **Fig. 1d**. The absorption exhibits a steep optical band edge with an excitonic feature at 2.32 eV, confirming the signature of polycrystalline $FAPbBr_3$ films consistent with previous studies.[15] As the concentration of additive increases up to 2%, new features arise below the bandgap of the host $FAPbBr_3$ perovskite, as evidenced by the peak at 1.6 eV in PL spectra (**Fig. 1d**, solid lines). We assign this 1.6 eV feature to the formation of I-rich phases $FAPb(I_{1-x}Br_x)_3$ in the host matrix. We further investigated the microscopic origin of the emission signatures in the films using wide-field PL imaging (**Fig. 1e**), which reveals the presence of red-color emitters against green-emitting $FAPbBr_3$ background at 2.28 eV. These site-specific emitting centers are more clearly visible with a 633 nm long-pass filter blocking the host PL (**Fig. 1f**). The red-emitting spots appear diffraction-limited under confocal microscope (~0.4 μm spatial resolution, **Supplementary Fig. 2a-2b**), indicating site-specific I-rich domains in the $FAPbBr_3$. Photoluminescence excitation (PLE) spectra (**Supplementary Fig. 2c**) of the emitting centers further reveal that these features can be directly photo-excited below the bandgap of the $FAPbBr_3$, thus implying that they do not arise from traps of interface defects when heterostructures are created between semiconductors.[16,17] Therefore, we attribute these PL signatures to the localized emission from $FAPb(I_{1-x}Br_x)_3$ b-PQDs with lower bandgap. Interestingly, the PL intensity reaches the maximum when excitation is resonant to the band-to-band transition of the host $FAPbBr_3$.This indicates efficient carrier transfer from the host matrix to the localized emitting center, consistent with a nested, type-I energy band alignment that would arise from $FAPbI_3$ dots in a $FAPbBr_3$ matrix (**Fig. 1g**).[18] In addition, we detect a red-shifted band gap of the host matrix in the high doping regime (~5%) (**Supplementary Fig. 2d-g**), indicating that a higher fraction of excess iodine introduced by additive with bromide in the $FAPbBr_3$ lattice, as widely reported in the mixed-halide perovskite films.[19] We find that 1-2% of doping gives the optimal density of the emitting centers, without introducing unwanted halide mixture within the host perovskite.

Here, we propose that the b-PQD growth in the host matrix is dominated by a temperature-assisted LaMer-type and Ostwald ripening process.[7,20] As displayed in **Fig. 1h**, we hypothesize that the nucleation and growth of b-PQDs occur because of the presence of iodoplumbate ($[PbI_6]^{4-}$) complexes in the precursor solvents containing the host precursors (FABr and $PbBr_2$) in DMF/DMSO. Dynamic light scattering (DLS) measurements on precursor solution containing the 2% additive showed a monodispersed distribution with an average size less than 1 nm, consistent with the size of molecular iodoplumbate complexes (**Supplementary Fig. 6**). The films were then spin-coated at ambient conditions to remove the excess solvent and were flash annealed by transferring them onto a hot plate, which results in rapid LaMer-type nucleation where FA was consumed from the host to dominantly form $FAPbI_3$, which was sequentially followed by the formation of $FAPbBr_3$ host matrix. We believe that the nucleation sequences result from the drastic difference in the solubility behavior between $FAPbI_3$ and $FAPbBr_3$ at the high temperature regime, where the solubility of $FAPbI_3$ decreases significantly with increasing temperature up to 120 °C.[21] Indeed, the soft Lewis acid nature of $Pb^{2+}$ prefers the coordination with soft $I^-$ Lewis base ligand (over the harder $Br^-$ ligand) to form more stable I-rich complexes ($PbI_4^{2-}$, $PbI_3^-$, …). This tendency drives the preferred formation of I-rich $FAPbI_{3-x}Br_x$ species during annealing, as these iodine-based complexes exhibit greater strength and stability.[22,23] The subsequent cooling of the films to room temperature locks the b-PQDs in place. We also demonstrated that our approach for the growth of b-PQDs using flash annealing for the growth of b-PQDs applies to other perovskite systems such as $FAPbCl_3$ host with either $FAPbI_3$ or $FAPbBr_3$ b-PQDs using the corresponding iodide and bromide-based additives (**Supplementary Fig. 7f**).

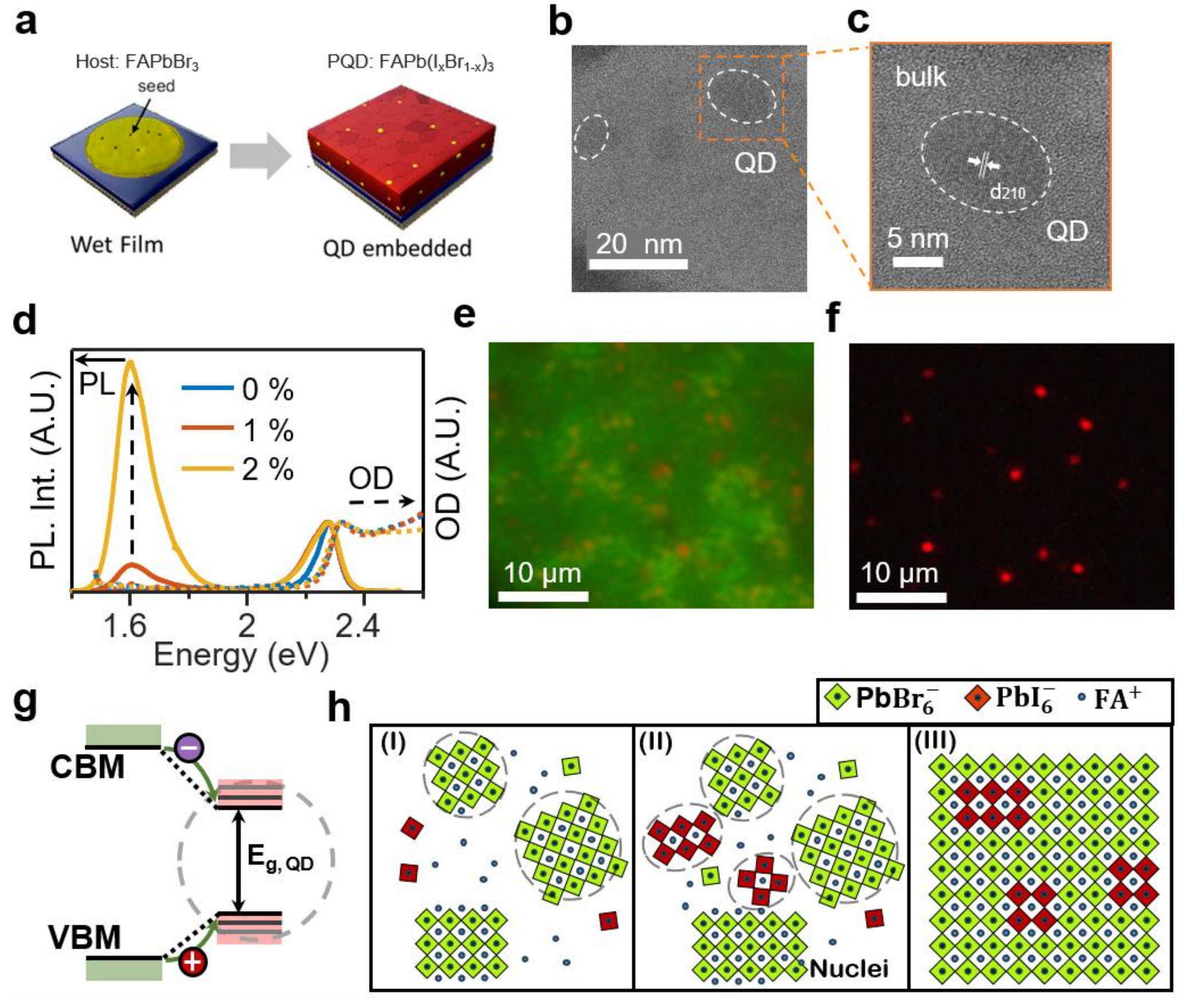

**Fig. 1 | Growth of buried quantum dots (b-PQDs) in a perovskite film**. **(a)** Synthesis of b-PQDs system using a one-step method. **(b)** Cross-section TEM images of PQD in matrix showing the presence of nanostructures. **(c)** Zoomed in view of orange-boxed regions in **(b)** showing a single b-PQD. The lattice fringes are indexed to (210) planes of $FAPb(I_{1-x}Br_x)_3$. **(d)** Ensemble optical absorbance and PL of b-PQDs in the $FAPbBr_3$ film. **(e)(f)** Wide-field PL imaging of b-PQDs within the same region of interest, with **(e)** 532nm and **(f)** 633nm long-pass filter, respectively. The filtered green and red regions indicate background emission from $FAPbBr_3$ matrix and localized emitting centers from b-PQDs. **(g)** Energy band alignment diagram indicating carrier transfer from $FAPbBr_3$ perovskite matrix to b-PQDs. **(h)** Formation kinetics of b-PQDs in $FAPbBr_3$ perovskite matrix.

## Photo-physics and quantum emission from individual QDs.

The buried quantum dots platform facilitates the formation of monodispersed individual PQDs within the perovskites film, serving as efficient and bright emiters of single photons. To explore the photo-physics of light emission from the b-PQD, we intensively studied the PL spectrum of the individual b-PQDs at cryogenic temperatures. At low-temperature (T ~ 6 K), inhomogeneous broadening due to phonon interactions are greatly reduced, and the PL of a single emitter exhibits a sharp emission peak (zero-phonon line, ZPL) resulting from direct exciton recombination. **Fig. 2a** shows PL spectra for 4 different b-PQDs in a $FAPbBr_3$ film, with near resonant excitation of 633 nm (1.96 eV), below the bandgap of host $FAPbBr_3$ (2.2 eV) at 4K.[25] With 2% doping, we detect the characteristic emission signatures from four b-PQDs, with a bright and narrow emission line at 1.922, 1.872, 1.818, and 1.722 eV, respectively. The existence of local quantum emitters has been previously reported in $MAPbI_3$ microplates and $CsPbBr_3$ microcrystals, originated by the localized exciton emission from local energy potential due to thickness variations.[26,27] Here, we note that these sharp emissions in our system are not detected from pristine $FAPbBr_3$ under the same excitation condition (**Supplementary Fig. 7a**), indicating the new emission states are introduced by the additive, instead of the intrinsic effects in $FAPbBr_3$.[16] **Fig. 2b** shows the PL and traces of single b-PQD (dot 4), with collected photon count rates of $10^3$ counts per second at low excitation intensity of 1 W/cm$^2$ at 633 nm under a high numerical aperture (NA = 0.95) objective. Such detected photon rates give comparable emission intensity as colloidal nanocrystals such as $CsPbBr_3$, $CsPbI_3$ and $FAPbI_3$, consistent with near-unity quantum yield at cryogenic temperature (**Supplementary Discussion 1**).[28–31] Furthermore, the PL intensity traces (**Fig. 2b & 2c**) show no signs of blinking, a key factor in reducing the emission intermittency in colloidal quantum dots, crucial for stable single-photon sources for reliable quantum applications.[6,32] Moreover, the b-PQDs also exhibit minimal spectral diffusion of less than 250 μeV with excellent photo stability (**Supplementary Fig. 8**). To quantify the spectral diffusion across the entire film, we measured the average spectral diffusion across tens of measured dots per film, which resulted in a spectra diffusion of less than 1 meV (**Supplementary Fig. 8b**), indicating excellent stability of b-PQDs desired for applications. These spectral trajectories show strong spectral stability among the FA-based hybrid c-PQDs ($FAPbBr_3$ and $FAPbI_3$), but in comparison all-inorganic c-PQDs $CsPbI_3$ exhibit better PL stability[33], suggesting that the random organic cation dipole organizations (FA reorientation) in both b-PQDs and c-PQDs systems may play a critical role for the spectral diffusion of the exciton emission line.[29,34]

To study the exciton-phonon coupling in a single b-PQDs, we investigated the phonon sidebands of exciton emission lines as well as the temperature dependence of the exciton linewidth. A zero-phonon line (ZPL) and phonon replica of dot 3 are shown in **Fig. 2d**. Three photon replicas can be detected, offset from the ZPL by 3-4 meV, 6-7 meV and 10 meV, respectively. These sidebands are assigned to different LO phonon modes, whose phonon energies lie in the same energy range as observed in the low temperature spectroscopy from single $FAPbI_3$ and $FAPbBr_3$ colloidal PQDs, as well as for both the bulk single crystals studied at low temperature by neutron scattering.[29,35,36] These detected phonon side-bands originate from the stretching and bending modes of $PbI_3$ counterpart and mutually-coupled motion of FA cations.[29] Furthermore, the evolution of PL spectra as a function of temperature as shown in **Fig. 2e**, reveals the PL from a single b-PQD, which experiences a systematic blue shift and thermal broadening as the temperature is increased from 3.7 K to 120 K. Such bandgap widening of embedded PQDs with increasing temperature is widely reported in bulk lead halide perovskites, attributed to reduced antibonding overlap between Pb-6s and I-5p orbitals, increasing the overall bandgap.[25,37] As demonstrated in **Fig. 2f**, the temperature dependent linewidth exhibits a weak acoustic phonon contribution to the broadening (exciton–acoustic phonon coupling of $\sigma_{Ac} < 10$ μeV $K^{-1}$), and exciton-phonon coupling with effective LO phonon energy of  20 meV, consistent with the reported study on $FAPbI_3$ PQDs and thin films.[25,29]

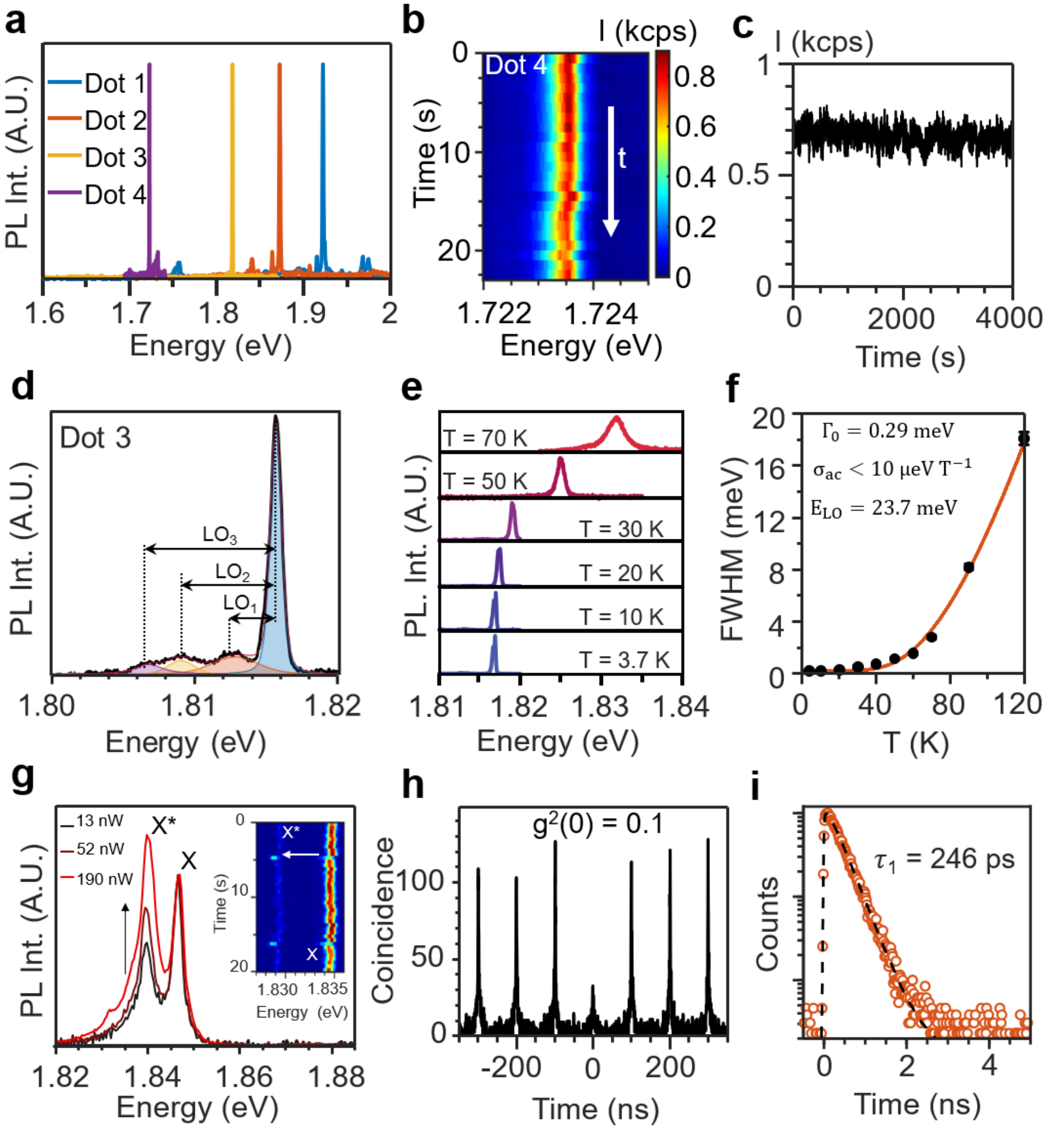

**Fig. 2 | Optical characteristics and intrinsic exciton linewidth of single b-PQDs. (a)** Typical PL spectra of four b-PQDs measured at 4K, showing narrow linewidths with peak emission from 1.7-2.1 eV. **(b)** PL traces of dot 4 over 25s with 1s binning, showing minimal blinking and spectral diffusion of ~250 μeV. **(c)** Long-time intensity trace of dot 5 with 1 s binning, with blinking-free signature for 4000s. **(d)** Zero-phonon line and phonon side band of neutral exciton PL in dot 3. **(e)** Temperature-dependent PL of dot 3. **(f)** PL broadening of dot 3 showing exciton-LO phonon coupling characters. **(g)** PL spectrum under high excitation. Multiple emission lines (such as trion X*) are detected with increasing excitation. Inset: PL traces indicating anticorrelation between X and X* due to photo-charging. **(h)** $2^{nd}$ order intensity correlation ($g^2(\tau)$) of dot 5, where photon-antibunching of $g^2(0) = 0.1$ is observed. **(i)** The exciton decay dynamics of dot 5, following mono-exponential decay with 246 ps lifetime.

## Exciton and trion fine-structure in b-PQDs

Because of the spatial confinement and enhanced Coulomb interactions between electrons and holes in PQDs, the formation of multi-exciton states, such as biexcitons and trions, is more likely in PQDs than in bulk semiconductors.[38] As indicated in **Fig. 2g**, the PL spectra of the individual b-PQD reveal multiple emission lines, whose responses change with the excitation density.[29] The splitting of the emission lines is detected at 1.845, 1.840 eV separately. We attribute the emission states to the neutral exciton (X) and the trion state (X*). As shown in **Fig. 2g**, the trion state demonstrates a superlinear dependence and dominates over the exciton at higher excitation density. The inset plots the PL time traces of single b-PQD, where a clear anticorrelation from X to X* emission lines is observed, indicating temporal energy transfer from X to X* state emission. Furthermore, the assignment of the fine-structure is further confirmed by direct spectroscopic signatures under a magnetic field at cryogenic temperatures.

At the single-dot level, the b-PQDs also exhibit quantum emission, evidenced by strong photon antibunching displayed in **Fig. 2h** measured on an isolated b-PQD and spectrally filtering the mono excitonic state. We found auto-correlation function $g^2(0) \sim 0.1$, which is consistent with the single photon emission of the embedded emitting centers, confirming that b-PQDs remarkably maintain the quantum nature of light even when buried in a polycrystalline film. The non-zero value of $g^2(0)$, despite the ultrasharp linewidth, can be attributed to the residual excitation of exciton complexes or simultaneous excitation of additional emission sites adjacent to the b-PQDs. As shown in **Fig. 2i**, the radiative exciton decay extracted from time-resolved PL suggests a mono-exponential decay with a lifetime of 246 ps, originating from single exciton recombination.

Besides evidence of SPE, the b-PQDs exhibit rich photo-physics from exciton and trion fine structures that can be studied in detail, thanks to the intrinsic linewidth far below the one measured for pure $FAPbI_3$ and $FAPbBr_3$ colloidal PQDs.[29,35] To study these properties of fine-structure splitting (FSS) and multi-exciton emissions, we performed high-resolution single-dot PL studies with 120 μeV spectro resolution. As exemplified in **Fig. 3a**, the PL spectra of b-PQD dot 3 display an emission doublet (X, 1.8167 eV with zero-field splitting of 270 μeV), as well as a trion emission (X*, 1.8135eV) at the low energy side. The splitting of the excitonic emissions is attributed to the fine-structure of the band-edge exciton with bright triplet level degeneracy lifting induced by the reduced crystal symmetry or shape anisotropy of the b-PQDs.[35,39] The bright excitonic transition of a single b-PQD reveals sharp, resolution-limited emission lines (FWHM ~130 μeV, **Supplementary Fig. 9**), similar to those observed in colloidal PQDs at cryogenic temperatures. [28,29,33,40] In contrast, the homogeneous broadening of the trion remains inaccessible in $FAPbBr_3$ or $FAPbI_3$ films, where intrinsic linewidths are typically broader (15-20 meV) due to inhomogeneous broadening.[25]

To further characterize the optical properties of the exciton and trion states, we perform magneto-dependent PL up to 7T, which has been shown to lift the exciton state degeneracy and reveal the exciton fine-structure via brightening and Zeeman splitting. **Fig. 3b** displays the PL of same b-PQD with the presence of magnetic field (B = 6T), demonstrating a triplet splitting of neutral exciton state (X, 1.8174 eV, 1.8171 & 1.8165 eV), as well as a magnetic brightening of a singlet state arising at 1.8155 eV. In addition to the exciton FSS, a trion splitting into doublets (X*, **Fig. 3a &3b**) at the low energy side is revealed with increasing magnetic field, expected response as in c-PQDs.[39] These spectral behaviors are the typical signatures of the exciton FSS and dark singlet state brightening, consistent with previous magneto-optical studies on colloidal $FAPbI_3$ and $FAPbBr_3$ PQDs.[39] An energy level diagram describing the exciton and trion fine-structures of embedded PQDs is displayed in **Fig. 3c**. The band-edge exciton of perovskites is split by electron-hole exchange interaction into a dark ground singlet state (**D,** with zero angular momentum J = 0) and a bright triplet (**B,** with angular momentum unity J = 1). As shown in **Fig. 3d**, the PL of the single b-PQD demonstrates systematic Zeeman shift neutral exciton and dark state brightening, with increasing magnetic field up to 7 T. The extracted peak energy splitting within the four exciton emission lines is displayed in **Fig. 3e**, which is defined with respect to the average value of four fine-structure states ($E_i - \overline{E}$, i=1-4). The magnetic-field dependent energy splitting of the exciton triplets and trion doublets reveal a g factor of 2.6 (**Supplementary Fig. 10**), consistent with the reported and predicted magnetic-field effect on exciton photoluminescence.[39,41]

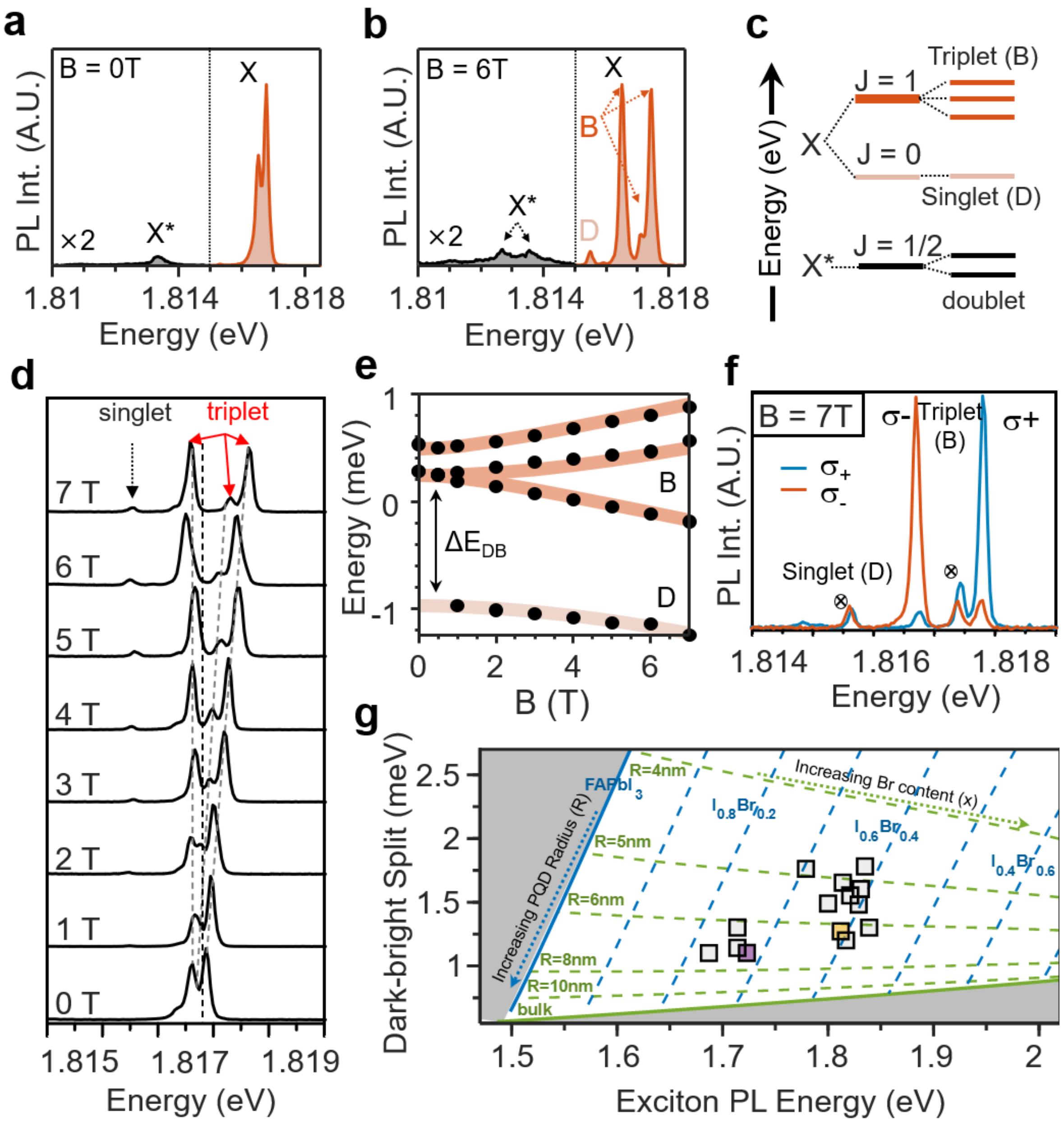

**Fig. 3 | Neutral and charged exciton fine-structures in b-PQDs. (a)** High-resolution PL spectrum of a single b-PQD (dot 3) at 4K, showing exciton features (X, solid red) and the trion state (X*, solid black). Spectra below 1.815eV are magnified (×2) for visualization. **(b)** PL spectrum of dot 3 under magnetic field B = 6T. The new features are assigned to bright triplets (B), dark singlet (D) and trion doublets (X*). **(c)** Schematics of fine-structures of exciton (X) and trion states (X*). **(d)** PL of b-PQD collected under magnetic field from 0-7T, revealing progressive evolution of triplets and singlet. **(e)** Extracted relative energy shift of exciton fine-structures with magnetic field. **(f)** Polarization dependence of the triplet states at 7T. The emission line at 1.8177 eV (1.8167 eV) shows σ+ (σ-) polarization, respectively. **(g)** Calculated exciton emission energy and dark-bright splitting of b-PQDs. The statistics of measured b-PQDs are indicated on the diagram PQD 3-4 are listed as matching colors in Fig. 2a.

Due to the lifted bright state degeneracy, the detected exciton fine-structure in b-PQDs exhibits a strong polarization dependence under a high magnetic field. As shown in **Fig. 3f**, the high and low energy lines of the exciton triplets show circularly polarized emissions with opposite helicity, where the high energy (1.8177 eV) reveals a dominant $\sigma^+$ polarization, and the low energy (1.8167 eV) exhibits $\sigma^-$ polarization, respectively. This polarization-dependent PL agrees with the eigenstates assignment under a magnetic field, as sketched in the energy diagram in **Fig. 3c**. We note that the central line and the brightened dark state line show no clear polarization orientation, which could be attributed to the transition dipole moment parallel to the magnetic field in the Faraday configuration.[30,42] These polarization behaviors of the fine-structures further confirm the excitonic nature of the b-PQDs.

To understand the nature of the emission line distributions and their relationship to the size and composition of the b-PQDs, we analyze the PL spectra and dark-bright splitting based on the simulation of excitonic states using finite confinement potentials between the b-PQDs and the wide-bandgap host for both electrons and holes. The calculations predict that embedded PQDs exhibit both size and composition dependent exciton emission energy and dark-bright splitting of the excitonic emission lines, consistent with the size and composition distribution from TEM analysis (**Supplementary Fig. 1**). To reproduce and quantify these relationships, we calculate the exciton emission energy and exciton fine-structures for different b-PQD radius and Br/I ratio. **Fig. 3g** gives a direct phase diagram of the predicted PL emission and dark-bright splitting of the PQDs in perovskite as a function of the size and chemical compositions ($0<x<1$: Br content). Within the same chemical composition (x), both exciton energy and energy splitting display an increase with reduced PQD sizes (blue lines in **Fig. 3g**), as a result of size-dependent exciton quantum confinement and long-range contribution of the exchange interactions to the energy splitting, consistent with the reported scaling laws for perovskite nanocrystals.[39] The typical observed exciton emissions of b-PQDs fall in the grided unshaded area, suggesting a distribution of PQD sizes and mixed bromide iodide content with $0.2<x_{Br}<0.4$. The dot 3 and dot 4 from **Fig. 2a** are marked on the same plot with matching colors, indicating estimated PQD composition and size (d~10nm) based on the statistical response from TEM characterizations.

## Electrically driven single-photon sources based on b-PQDs

The photo-physics of the dot in perovskite film suggests the preservation of the quantum nature of light with on-demand emission. The success of embedding the emitters into perovskite film gives possibilities, including electrically driven SPEs, which has remaineda major challenge for c-PQDs passivated with insulating ligands. In order to demonstrate the electrically injected quantum emission from b-PQDs, we fabricated a device using a standard LED-like architecture of ITO / PEDOT:PSS / PVK / $FAPbBr_3$ (SPE embedded, thickness of ~100 nm) / TPBi / LiF / Au, as indicated in **Fig. 4a**.[43] We investigate the response of the quantum light LED under both room temperature and cryogenic temperatures, as a direct correlation with the photo-physical behaviors in **Fig. 2** and **Fig. 3**. As indicated in **Fig. 4b**, the current-voltage response of a single device exhibits a diode-like behavior under 295 K and 6 K. The forward bias of the devices shows an increase from 3 V to 7 V with decreasing temperature, suggesting the increase of series resistance originated from the reduced conductivity due to the lower population of thermally-activated carriers.[44] The device shows strong green emission under 7 V, primarily from the carrier recombination in the host matrix (2.2 eV). Interestingly, as displayed in **Fig. 4c** the perovskite active layer demonstrates a non-uniform emission from localized emitting centers (**Fig. 4c** in red). Like the optical response, the EL spectra show two features: the band edge transition from $FAPbBr_3$ (2.2 eV) and a narrow low energy emission state (2.015 eV), which resembles the PL indicated in **Fig. 3a**. As demonstrated in **Fig. 4e**, the embedded emitter shows clear energy splitting, assigned to the biexciton transition, evidenced by the quadratic and linear dependence of the injection current density (**Fig. 4f**). This may be indicative of a more balanced charge injection to the b-PQD, than with optical pumping. **Fig. 4g** monitors the $g^2(\tau)$ of the same spot in **Fig. 4c**, which shows clear photon antibunching of $g^2(0) = 0.44$ (**Fig. 2i**). As suggested in **Fig. 4h**, the kinetics of the EL spectra demonstrate blinking and diffusion behavior, which may be attributed to the trion states. In addition, the biexciton emission contributes to the non-zero $g^2(0)$ observed in **Fig. 4g**. The clear photon-antibunching from electrical luminescence suggests the fact that the emitters in the perovskite matrix can be both optically and electrically pumped, paving the way for future application for ideal practical single photon sources.

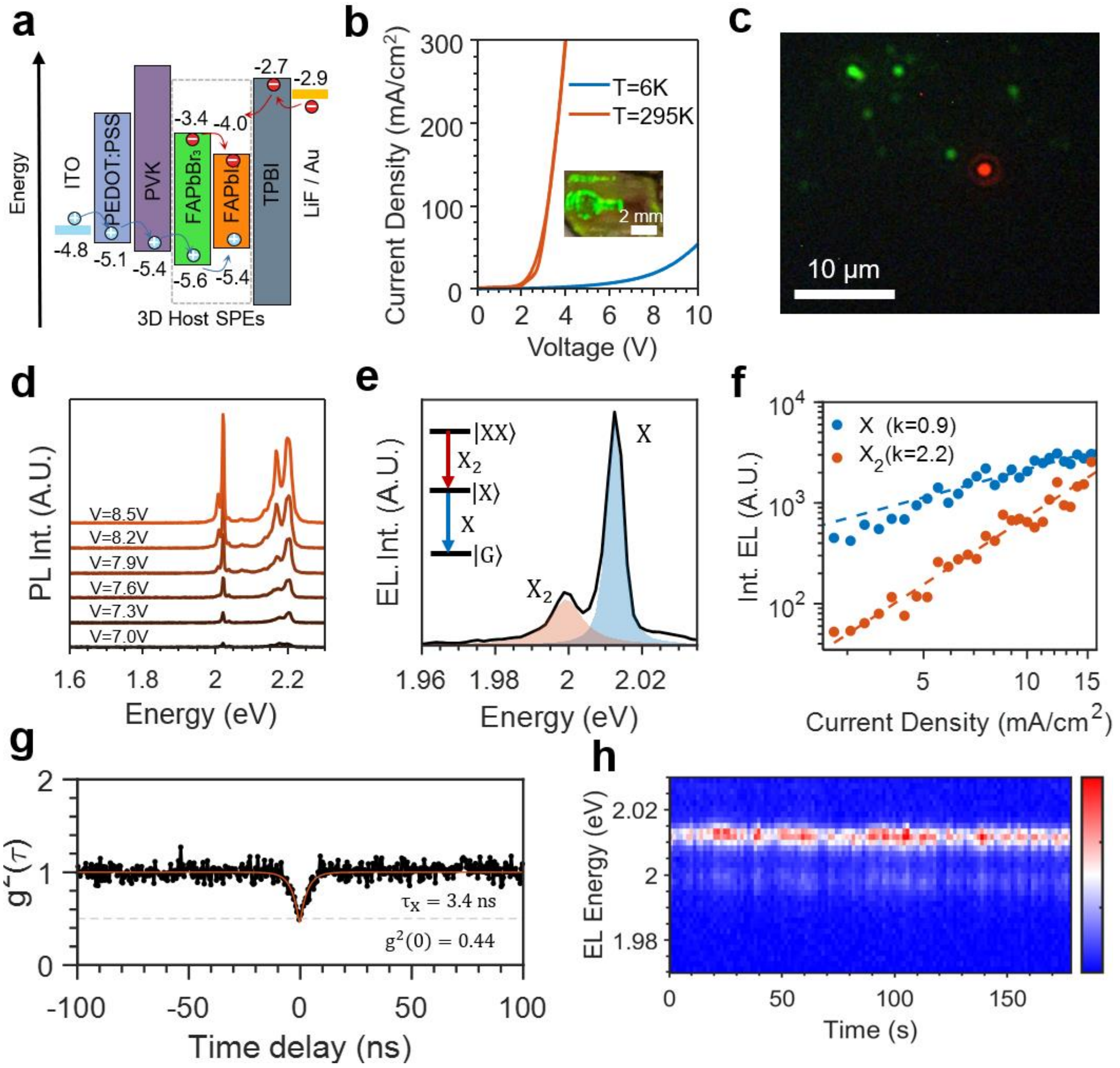

**Fig. 4 | Electrically – driven single photon emission based on b-PQDs. (a)** Device architecture of b-PQDs and energy band alignment of each layer. **(b)** Current density - voltage curves of the PQD-in-matrix system at T = 295 K and T = 6 K. **(c)** Electro-luminescence (EL) mapping at T=6K showing diffraction-limited localized emitting centers. **(d)** EL of b-PQDs with increasing voltage bias. **(e)** splitting of emission lines assigned to exciton and bi-exciton states. **(f)** EL Intensity as function of current revealing linear and quadratic dependence of X and X2 states. **(g)** Electrically-driven single-photon emission of single dot in **(c)** at bias of 9.9 V (I=0.59 mA), showing antibunching with $g^2(0) = 0.44$. **(h)** Time series showing blinking behavior of the b-PQDs under electrical excitation.

# Methods

## Materials

The $PbBr_2$ (>98%) and $PbI_2$ (>98%) used for the dot-in-matrix synthesis were purchased from TCI. FABr (>99.99%) was purchased from GreatCell Solar. For devices, Poly(9-vinylcarbazole) (PVK), LiF (>99.99%) and 2,2',2"-(1,3,5-Benzinetriyl)-tris(1-phenyl-1-H-benzimidazole) (TPBi) were purchased from Sigma Aldrich, poly(3,4-ethylenedioxythiophene) polystyrene sulfonate (PEDOT:PSS) was purchased from Heraeus, and Au was purchased from Kurt J. Lesker.

## Device fabrication for electrical SPEs

The device was fabricated using a LED-like architecture of ITO / PEDOT:PSS / PVK / $FAPbBr_3$ (SPE embedded) / TPBi / LiF / Au. Patterned glass/ITO was cleaned following the procedure above. PEDOT:PSS was spin-coated outside a glovebox at 2000 RPM for 30 sec, followed by annealing at 120 $^oC$ for 30 min. PVK was dissolved in chlorobenzene (6mg/mL) and dynamically spin-coated in a glovebox at 4000RPM for 30 sec, followed by annealing at 170 $^oC$ for 30 min. The perovskite layer was deposited as above. TPBi (35 nm), LiF (1 nm), and Au (80 nm) were deposited by thermal evaporation under a vacuum of less than $2 \times 10^{-6}$ torr. The active area selected for the devices was 3.1 $mm^2$.

## Optical spectroscopy measurements

**Absorbance, PL, and low-temperature spectroscopy.** The optical absorbance of the samples was measured by illuminating the samples using a broadband light source (Thorlabs SOLIS–3C), which was focused on the samples through an objective lens yielding a 50 μm spot size. The absorbance spectrum was collected based on a transmission geometry. The PL spectroscopy was characterized based on a lab-built confocal microscopic system. A monochromatic pulsed laser (repetition rate 78.1 MHz, NKT Photonics) was used to photo-excite the samples. The excitation wavelength was 480 nm for above-bandgap excitation and 590 nm for below-bandgap excitation, unless mentioned otherwise. The laser was focused onto the sample through an objective lens (50×, 0.42 NA, Mitutoyo) with ~ 1μm beam size, yielding an average excitation intensity of $3\times10^2$ $W/cm^2$. The PL data was collected by a spectrometer (Kymera 328i, Andor) and a CCD camera (iDus 416, Andor). For cryogenic PL measurements, the sample was kept under vacuum ($10^{-4}$ to $10^{-5}$ torr) in a closed-cycle cryostat (Advanced Research Systems) and maintained at cryogenic temperature (T = 6.5K).

**PL imaging.** For wide-field PL imaging of the buried PQD films, a light-source was filtered by short-pass filter (BSP532) before incidence on the sample. The photoluminescence of the sample was spectrally selected using a dichroic beamsplitter (DBS532) and a long-pass filter (LP532) before being imaged by the camera. The spatial resolution is 1μm calibrated by the dispersed colloidal single PQDs. For imaging of the embedded PQDs, the PL was further filtered using LP633 to isolate the narrow emission of the embedded PQDs. For confocal PL imaging, 20 μm by 20 μm region of interest was measured with a step size of 0.5 μm.

**Time-correlated single-photon counting.** Time-resolved PL was collected using the 590nm NKT Photonics laser at 26 MHz repetition, and photon counts were acquired using a single-photon avalanched photodiode (SAPD) and Picoquant Hydraharp 400. For photon-antibunching measurements, two SPADs were used and set up in Hanbury Brown and Twiss (HBT) geometry. The narrow emission was spectrally filtered by the spectrometer before being detected by the SPAD detectors. The photon events were collected by Picoquant hydraharp 400 and analyzed using the Python library readPTU.[45]

**PQD magneto-spectroscopy.** A home-built confocal microscope based on a NA=0.95 objective was placed in a cryostat and used to characterize the sample. The samples were excited using a 633 nm pulsed laser with 80 MHz repetition rate. The spectro-resolution is 120 μeV using 1800 lines/mm grating. The magneto-optical properties of the PQDs are explored in the Faraday configuration, where the applied field (up to 7T) is parallel to the microscope optical axis.

**Electrical-luminescence spectroscopy.** The fabricated device was loaded into a vacuum optical cryostat (Montana Instruments, ~$5\times10^{-8}$ Torr) for electrical-luminescence measurements. All measurements are performed at 30K. The voltage bias is applied by a Keithley 2400 source meter. The emission photons are collected by a 100× objective with NA of 0.7 and then guided to Horiba iHR 320 CCD spectrometer through free space optics based on a home-built Raman spectroscopy setup. The spectral resolution was approximately 0.5 nm, and the spatial resolution, controlled by the width of the spectrometer's entrance slit, was around 1 μm, capturing only the red emitter as shown in Fig. 4c. After obtaining a satisfactory spectrum, time-correlated single-photon counting measurements were performed.

## Simulation of buried PQD emission energy

The buried PQDs were modeled as a sphere of radius R composed of $FAPb(I_{1-x}Br_x)_3$ inside a $FAPbBr_3$ matrix. The electron-hole excitation was modeled using an effective mass framework, with finite confinement barriers defined by the band alignments between the PQD and the surrounding matrix.[46,47] The parameters for the bulk electronic structure of $FAPb(I_{1-x}Br_x)_3$ were estimated from those available for $FAPbI_3$ and $FAPbBr_3$ using Vegard's law.[48–51] The exciton wavefunction and energy were found through a standard variational ansatz and the dark-bright splitting was subsequently calculated using a perturbative evaluation of the electron-hole exchange interaction.[35,52,53] The formalism and material parameters are further detailed in the Supplementary Information available online.

## GIWAXS measurements

The GIWAXS measurements presented in this paper were conducted at 11-BM at the National Synchrotron Light Source-II (NSLS II). Experiments at beamline 11-BM utilized a robotic stage within a vacuum chamber maintained at a pressure of $6 \times 10^{-2}$ torr. A Pilatus 1M (Dectris) area detector was situated approximately 267 mm away from the sample was employed. The photon energy used was 13.5 keV, and the X-ray beam had dimensions of 200 µm × 50 µm (horizontal × vertical).

## TEM measurements

TEM samples were prepared via focused ion beam (FIB) milling to identify the distribution of the quantum dots. Cross-sectional TEM samples prepared via Focused Ion Beam (FIB) milling for identifying the distribution of the quantum dots. The low dose TEM experiments were performed using the Titan microscope at 300 kV installed with a direct electron detector (Gatan K3) with a dose rate of 2000 $e/Å^2$/sec and 0.5-1s exposure time.

## References

1. Akkerman, Q. A., Rainò, G., Kovalenko, M. V. & Manna, L. Genesis, challenges and opportunities for colloidal lead halide perovskite nanocrystals. *Nature Mater* **17,** 394–405 (2018).
2. Dey, A., Ye, J., De, A., Debroye, E., Ha, S. K., Bladt, E., Kshirsagar, A. S., Wang, Z., Yin, J., Wang, Y., Quan, L. N., Yan, F., Gao, M., Li, X., Shamsi, J., Debnath, T., Cao, M., Scheel, M. A., Kumar, S., Steele, J. A., Gerhard, M., Chouhan, L., Xu, K., Wu, X., Li, Y., Zhang, Y., Dutta, A., Han, C., Vincon, I., Rogach, A. L., Nag, A., Samanta, A., Korgel, B. A., Shih, C.-J., Gamelin, D. R., Son, D. H., Zeng, H., Zhong, H., Sun, H., Demir, H. V., Scheblykin, I. G., Mora-Seró, I., Stolarczyk, J. K., Zhang, J. Z., Feldmann, J., Hofkens, J., Luther, J. M., Pérez-Prieto, J., Li, L., Manna, L., Bodnarchuk, M. I., Kovalenko, M. V., Roeffaers, M. B. J., Pradhan, N., Mohammed, O. F., Bakr, O. M., Yang, P., Müller-Buschbaum, P., Kamat, P. V., Bao, Q., Zhang, Q., Krahne, R., Galian, R. E., Stranks, S. D., Bals, S., Biju, V., Tisdale, W. A., Yan, Y., Hoye, R. L. Z. & Polavarapu, L. State of the Art and Prospects for Halide Perovskite Nanocrystals. *ACS Nano* **15,** 10775–10981 (2021).
3. Somaschi, N., Giesz, V., De Santis, L., Loredo, J. C., Almeida, M. P., Hornecker, G., Portalupi, S. L., Grange, T., Antón, C., Demory, J., Gómez, C., Sagnes, I., Lanzillotti-Kimura, N. D., Lemaítre, A., Auffeves, A., White, A. G., Lanco, L. & Senellart, P. Near-optimal single-photon sources in the solid state. *Nature Photon* **10,** 340–345 (2016).
4. Borri, P., Langbein, W., Woggon, U., Stavarache, V., Reuter, D. & Wieck, A. D. Exciton dephasing via phonon interactions in InAs quantum dots: Dependence on quantum confinement. *Phys. Rev. B* **71,** 115328 (2005).
5. Efros, A. L. & Nesbitt, D. J. Origin and control of blinking in quantum dots. *Nature nanotechnology* **11,** 661–671 (2016).
6. Nguyen, H. A., Dixon, G., Dou, F. Y., Gallagher, S., Gibbs, S., Ladd, D. M., Marino, E., Ondry, J. C., Shanahan, J. P., Vasileiadou, E. S., Barlow, S., Gamelin, D. R., Ginger, D. S., Jonas, D. M., Kanatzidis,

M. G., Marder, S. R., Morton, D., Murray, C. B., Owen, J. S., Talapin, D. V., Toney, M. F. & Cossairt, B. M. Design Rules for Obtaining Narrow Luminescence from Semiconductors Made in Solution. *Chem. Rev.* **123,** 7890–7952 (2023).

7. Akkerman, Q. A., Nguyen, T. P. T., Boehme, S. C., Montanarella, F., Dirin, D. N., Wechsler, P., Beiglböck, F., Rainò, G., Erni, R., Katan, C., Even, J. & Kovalenko, M. V. Controlling the nucleation and growth kinetics of lead halide perovskite quantum dots. *Science* **377,** 1406–1412 (2022).

8. Almeida, G., Infante, I. & Manna, L. Resurfacing halide perovskite nanocrystals. *Science* **364,** 833–834 (2019).

9. Morad, V., Stelmakh, A., Svyrydenko, M., Feld, L. G., Boehme, S. C., Aebli, M., Affolter, J., Kaul, C. J., Schrenker, N. J., Bals, S., Sahin, Y., Dirin, D. N., Cherniukh, I., Raino, G., Baumketner, A. & Kovalenko, M. V. Designer phospholipid capping ligands for soft metal halide nanocrystals. *Nature* **626,** 542–548 (2024).

10. Ahmed, G. H., Yin, J., Bakr, O. M. & Mohammed, O. F. Successes and Challenges of Core/Shell Lead Halide Perovskite Nanocrystals. *ACS Energy Lett.* **6,** 1340–1357 (2021).

11. Talapin, D. V., Lee, J.-S., Kovalenko, M. V. & Shevchenko, E. V. Prospects of Colloidal Nanocrystals for Electronic and Optoelectronic Applications. *Chem. Rev.* **110,** 389–458 (2010).

12. Govinda, S., Kore, B. P., Swain, D., Hossain, A., De, C., Guru Row, T. N. & Sarma, D. D. Critical Comparison of FAPbX3 and MAPbX3 (X = Br and Cl): How Do They Differ? *J. Phys. Chem. C* **122,** 13758–13766 (2018).

13. Stoumpos, C. C., Malliakas, C. D. & Kanatzidis, M. G. Semiconducting Tin and Lead Iodide Perovskites with Organic Cations: Phase Transitions, High Mobilities, and Near-Infrared Photoluminescent Properties. *Inorg. Chem.* **52,** 9019–9038 (2013).

14. Huang, Y., Li, L., Liu, Z., Jiao, H., He, Y., Wang, X., Zhu, R., Wang, D., Sun, J., Chen, Q. & Zhou, H. The intrinsic properties of FA(1−x)MAxPbI3 perovskite single crystals. *J. Mater. Chem. A* **5,** 8537–8544 (2017).

15. Hanusch, F. C., Wiesenmayer, E., Mankel, E., Binek, A., Angloher, P., Fraunhofer, C., Giesbrecht, N., Feckl, J. M., Jaegermann, W., Johrendt, D., Bein, T. & Docampo, P. Efficient Planar Heterojunction Perovskite Solar Cells Based on Formamidinium Lead Bromide. *J. Phys. Chem. Lett.* **5,** 2791–2795 (2014).

16. Diab, H., Trippé-Allard, G., Lédée, F., Jemli, K., Vilar, C., Bouchez, G., Jacques, V. L. R., Tejeda, A., Even, J., Lauret, J.-S., Deleporte, E. & Garrot, D. Narrow Linewidth Excitonic Emission in Organic–Inorganic Lead Iodide Perovskite Single Crystals. *J. Phys. Chem. Lett.* **7,** 5093–5100 (2016).

17. Smith, M. D., Jaffe, A., Dohner, E. R., Lindenberg, A. M. & Karunadasa, H. I. Structural origins of broadband emission from layered Pb–Br hybrid perovskites. *Chem. Sci.* **8,** 4497–4504 (2017).

18. Ning, Z., Gong, X., Comin, R., Walters, G., Fan, F., Voznyy, O., Yassitepe, E., Buin, A., Hoogland, S. & Sargent, E. H. Quantum-dot-in-perovskite solids. *Nature* **523,** 324–328 (2015).

19. McMeekin, D. P., Sadoughi, G., Rehman, W., Eperon, G. E., Saliba, M., Hörantner, M. T., Haghighirad, A., Sakai, N., Korte, L., Rech, B., Johnston, M. B., Herz, L. M. & Snaith, H. J. A mixed-cation lead mixed-halide perovskite absorber for tandem solar cells. *Science* **351,** 151–155 (2016).

20. Carey, G. H., Abdelhady, A. L., Ning, Z., Thon, S. M., Bakr, O. M. & Sargent, E. H. Colloidal Quantum Dot Solar Cells. *Chem. Rev.* **115,** 12732–12763 (2015).

21. Petrov, A. A., Ordinartsev, A. A., Fateev, S. A., Goodilin, E. A. & Tarasov, A. B. Solubility of Hybrid Halide Perovskites in DMF and DMSO. *Molecules* **26,** 7541 (2021).

22. Kerner, R. A., Schloemer, T. H., Schulz, P., Berry, J. J., Schwartz, J., Sellinger, A. & Rand, B. P. Amine additive reactions induced by the soft Lewis acidity of Pb2+ in halide perovskites. Part I: evidence for Pb–alkylamide formation. *J. Mater. Chem. C* **7,** 5251–5259 (2019).

23. Weller, D. M., Weller, M. T., Overton, T., Rourke, J. & Armstrong, F. *Inorganic Chemistry*. (OUP Oxford, 2014).

24. Sidhik, S., Metcalf, I., Li, W., Kodalle, T., Dolan, C. J., Khalili, M., Hou, J., Mandani, F., Torma, A., Zhang, H., Garai, R., Persaud, J., Marciel, A., Muro Puente, I. A., Reddy, G. N. M., Balvanz, A., Alam, M. A., Katan, C., Tsai, E., Ginger, D., Fenning, D. P., Kanatzidis, M. G., Sutter-Fella, C. M., Even, J. & Mohite, A. D. Two-dimensional perovskite templates for durable, efficient formamidinium perovskite solar cells. *Science* **384,** 1227–1235 (2024).

25. Wright, A. D., Verdi, C., Milot, R. L., Eperon, G. E., Pérez-Osorio, M. A., Snaith, H. J., Giustino, F., Johnston, M. B. & Herz, L. M. Electron–phonon coupling in hybrid lead halide perovskites. *Nature Communications* **7,** 11755 (2016).

26. Feng, S., Qin, Q., Han, X., Zhang, C., Wang, X., Yu, T. & Xiao, M. Universal Existence of Localized Single-Photon Emitters in the Perovskite Film of All-Inorganic CsPbBr3 Microcrystals. *Advanced Materials* **34,** 2106278 (2022).

27. Yuan, J., Zhou, D., Zhuang, C., Zhou, Y., Zhang, C., Wang, L., Xiao, M. & Wang, X. Single-Photon Emission from Single Microplate MAPbI3 Nanocrystals with Ultranarrow Photoluminescence Linewidths and Exciton Fine Structures. *Advanced Optical Materials* **10,** 2200606 (2022).

28. Utzat, H., Sun, W., Kaplan, A. E. K., Krieg, F., Ginterseder, M., Spokoyny, B., Klein, N. D., Shulenberger, K. E., Perkinson, C. F., Kovalenko, M. V. & Bawendi, M. G. Coherent single-photon emission from colloidal lead halide perovskite quantum dots. *Science* **363,** 1068–1072 (2019).

29. Fu, M., Tamarat, P., Trebbia, J.-B., Bodnarchuk, M. I., Kovalenko, M. V., Even, J. & Lounis, B. Unraveling exciton–phonon coupling in individual FAPbI3 nanocrystals emitting near-infrared single photons. *Nat Commun* **9,** 3318 (2018).

30. Fu, M., Tamarat, P., Huang, H., Even, J., Rogach, A. L. & Lounis, B. Neutral and Charged Exciton Fine Structure in Single Lead Halide Perovskite Nanocrystals Revealed by Magneto-optical Spectroscopy. *Nano Lett.* **17,** 2895–2901 (2017).

31. Tamarat, P., Hou, L., Trebbia, J.-B., Swarnkar, A., Biadala, L., Louyer, Y., Bodnarchuk, M. I., Kovalenko, M. V., Even, J. & Lounis, B. The dark exciton ground state promotes photon-pair emission in individual perovskite nanocrystals. *Nat Commun* **11,** 6001 (2020).

32. Rainò, G., Nedelcu, G., Protesescu, L., Bodnarchuk, M. I., Kovalenko, M. V., Mahrt, R. F. & Stöferle, T. Single Cesium Lead Halide Perovskite Nanocrystals at Low Temperature: Fast Single-Photon Emission, Reduced Blinking, and Exciton Fine Structure. *ACS Nano* **10,** 2485–2490 (2016).

33. Yin, C., Chen, L., Song, N., Lv, Y., Hu, F., Sun, C., Yu, W. W., Zhang, C., Wang, X., Zhang, Y. & Xiao, M. Bright-Exciton Fine-Structure Splittings in Single Perovskite Nanocrystals. *Phys. Rev. Lett.* **119,** 026401 (2017).

34. Zhao, X.-G., Dalpian, G. M., Wang, Z. & Zunger, A. Polymorphous nature of cubic halide perovskites. *Phys. Rev. B* **101,** 155137 (2020).

35. Tamarat, P., Bodnarchuk, M. I., Trebbia, J.-B., Erni, R., Kovalenko, M. V., Even, J. & Lounis, B. The ground exciton state of formamidinium lead bromide perovskite nanocrystals is a singlet dark state. *Nat. Mater.* **18,** 717–724 (2019).

36. Ferreira, A. C., Paofai, S., Létoublon, A., Ollivier, J., Raymond, S., Hehlen, B., Rufflé, B., Cordier, S., Katan, C., Even, J. & Bourges, P. Direct evidence of weakly dispersed and strongly anharmonic optical phonons in hybrid perovskites. *Commun Phys* **3,** 1–10 (2020).

37. Dar, M. I., Jacopin, G., Meloni, S., Mattoni, A., Arora, N., Boziki, A., Zakeeruddin, S. M., Rothlisberger, U. & Grätzel, M. Origin of unusual bandgap shift and dual emission in organic-inorganic lead halide perovskites. *Science Advances* **2,** e1601156 (2016).

38. Kagan, C. R., Bassett, L. C., Murray, C. B. & Thompson, S. M. Colloidal Quantum Dots as Platforms for Quantum Information Science. *Chem. Rev.* **121,** 3186–3233 (2021).

39. Tamarat, P., Prin, E., Berezovska, Y., Moskalenko, A., Nguyen, T. P. T., Xia, C., Hou, L., Trebbia, J.-B., Zacharias, M. & Pedesseau, L. Universal scaling laws for charge-carrier interactions with quantum confinement in lead-halide perovskites. *Nature Communications* **14,** 229 (2023).

40. Kaplan, A. E. K., Krajewska, C. J., Proppe, A. H., Sun, W., Sverko, T., Berkinsky, D. B., Utzat, H. & Bawendi, M. G. Hong–Ou–Mandel interference in colloidal CsPbBr3 perovskite nanocrystals. *Nat. Photon.* **17,** 775–780 (2023).

41. Yu, Z. G. Effective-mass model and magneto-optical properties in hybrid perovskites. *Sci Rep* **6,** 28576 (2016).

42. Htoon, H., Crooker, S. A., Furis, M., Jeong, S., Efros, Al. L. & Klimov, V. I. Anomalous Circular Polarization of Photoluminescence Spectra of Individual CdSe Nanocrystals in an Applied Magnetic Field. *Phys. Rev. Lett.* **102,** 017402 (2009).

43. Zhao, H., Chen, H., Bai, S., Kuang, C., Luo, X., Teng, P., Yin, C., Zeng, P., Hou, L., Yang, Y., Duan, L., Gao, F. & Liu, M. High-Brightness Perovskite Light-Emitting Diodes Based on FAPbBr3 Nanocrystals with Rationally Designed Aromatic Ligands. *ACS Energy Lett.* **6,** 2395–2403 (2021).

44. Cao, X. A. & LeBoeuf, S. F. Current and Temperature Dependent Characteristics of Deep-Ultraviolet Light-Emitting Diodes. *IEEE Transactions on Electron Devices* **54,** 3414–3417 (2007).

45. Ballesteros, G. C., Proux, R., Bonato, C. & Gerardot, B. D. readPTU: a python library to analyse time tagged time resolved data. *J. Inst.* **14,** T06011 (2019).

46. Brus, L. E. Electron–electron and electron-hole interactions in small semiconductor crystallites: The size dependence of the lowest excited electronic state. *The Journal of Chemical Physics* **80,** 4403–4409 (1984).

47. Kayanuma, Y. & Momiji, H. Incomplete confinement of electrons and holes in microcrystals. *Phys. Rev. B* **41,** 10261–10263 (1990).

48. Galkowski, K., Mitioglu, A., Miyata, A., Plochocka, P., Portugall, O., Eperon, G. E., Wang, J. T.-W., Stergiopoulos, T., Stranks, S. D., Snaith, H. J. & Nicholas, R. J. Determination of the exciton binding energy and effective masses for methylammonium and formamidinium lead tri-halide perovskite semiconductors. *Energy Environ. Sci.* **9,** 962–970 (2016).

49. Tao, S., Schmidt, I., Brocks, G., Jiang, J., Tranca, I., Meerholz, K. & Olthof, S. Absolute energy level positions in tin- and lead-based halide perovskites. *Nat Commun* **10,** 2560 (2019).

50. Nestoklon, M. O., Goupalov, S. V., Dzhioev, R. I., Ken, O. S., Korenev, V. L., Kusrayev, Yu. G., Sapega, V. F., de Weerd, C., Gomez, L., Gregorkiewicz, T., Lin, J., Suenaga, K., Fujiwara, Y., Matyushkin, L. B. & Yassievich, I. N. Optical orientation and alignment of excitons in ensembles of inorganic perovskite nanocrystals. *Phys. Rev. B* **97,** 235304 (2018).

51. Becker, M. A., Vaxenburg, R., Nedelcu, G., Sercel, P. C., Shabaev, A., Mehl, M. J., Michopoulos, J. G., Lambrakos, S. G., Bernstein, N., Lyons, J. L., Stöferle, T., Mahrt, R. F., Kovalenko, M. V., Norris, D. J., Rainò, G. & Efros, A. L. Bright triplet excitons in caesium lead halide perovskites. *Nature* **553,** 189–193 (2018).

52. Kayanuma, Y. Wannier exciton in microcrystals. *Solid State Communications* **59,** 405–408 (1986).

53. Pikus, G. E. & Bir, G. L. Exchange interaction in excitons in semiconductors. *Sov. Phys. JETP* **33,** 108–114 (1971).

## Data availability

The data for this study is available from the corresponding authors upon reasonable request.

## Code availability

The analysis code for this study is available from the corresponding authors upon reasonable request.

## Acknowledgments

This material is based upon work supported by the National Science Foundation Science and Technology Center (STC) for Integration of Modern Optoelectronic Materials on Demand (IMOD) under Grant No. DMR-2019444. Any opinions, findings, conclusions or recommendations expressed in this material are those of the author and do not necessarily reflect the views of the National Science Foundation. S.S acknowledges the support of Eyring Materials Center for the use of electron microscopy facilities. S.S. and S.P.R thank Fulton School of Engineering start up funds for the research support. S.S. and S.P.R also thank Piyush Haluai for the staff support in acquiring the TEM datasets. J.S.C. and X.M. acknowledge support from the Center for Molecular Quantum Transduction, an Energy Frontier Research Center funded by the U.S. Department of Energy, Office of Science, Office of Basic Energy Sciences, under award no. DE-SC0021314. This work was performed, in part, at the Center for Nanoscale Materials, a DOE Office of Science User Facility, and supported by the U.S. Department of Energy, Office of Science, Office of Basic Energy Sciences under Contract No. DE-AC02-06CH11357.

## Author contributions

A.D.M. and J.E. conceived and designed the experiment. A.P., I.M., F.Z. and S.Sidhik. synthesized the PQDs in films. H.Z. performed the optical spectroscopy characterizations. P.T. and B.L. performed the magneto-spectroscopy of the PQD in perovskite samples. H.Z. performed the data analysis with guidance from J.E., A.D.M. and M.G.B.; Y.Z., S.L. and K.S. performed the electrical-injected SPE measurements with guidance of D.N; X.M. performed and analyzed the photon-antibunching study. I.M. performed GIWAX characterizations. F.M. performed DLS measurements. S.P.R. and S.Susarla performed the TEM and EDX mapping. J.H. and X.S. performed the SEM and AFM characterizations. S.T. performed the modeling of PQD emission energies with guidance from J.E. and C.K; M.S. performed magneto photoluminescence spectroscopy with guidance from P.T. and B.L.; H.Z., I.M., F.Z. and F.M. studied the formation kinetics of b-PQDs with guidance from A.D.M, D.S.G, and M.G.K; H.Z. and A.D.M. wrote the manuscript with inputs from all the authors. All the authors read the manuscript and agreed to its contents, and all the data are reported in the main text and Supplementary Information.

## Competing interests

Rice University have filed the patent for the method entitled “BURIED QUANTUM DOTS, METHODS AND USES THEREOF.”